\documentclass[ reprint,
superscriptaddress,
 amsmath,amssymb,  
 aps,longbibliography,prapplied
]{revtex4-2}

\usepackage[utf8x]{inputenc}
\usepackage[T1]{fontenc}
\usepackage{graphicx}% Include figure files
\usepackage{bm}% bold math
\usepackage[colorlinks=true,
            linkcolor=blue,
            citecolor=blue,
            urlcolor=blue]{hyperref}% add hypertext capabilities
\usepackage[separate-uncertainty=true,list-units=repeat,multi-part-units=brackets]{siunitx}
\usepackage{float}
\usepackage{svg}
\usepackage{orcidlink}

\makeatletter

\newcommand{\subfiglabel}[2]{%
  \begingroup
  \edef\@currentlabel{\thefigure(#2)}\phantomsection\label{#1}%
    \edef\@currentlabel{(#2)}\label{#1@sub}%
    \edef\@currentlabel{#2}\label{#1@lett}%
  \endgroup
}

\newcommand{\figsubref}[2]{%
  Fig.~\ref{#1}(%
  \begingroup
    \def\sep{}%
    \@for\sf:=#2\do{%
      \sep\hyperref[\sf]{\ref{\sf @lett}}%
      \def\sep{,}%
    }%
  \endgroup
  )%
}
\makeatother

\begin{document}

\preprint{APS/123-QED}
\pagestyle{plain}

\title{Fabrication, characterization and mechanical loading of Si/SiGe membranes\\ for spin qubit devices }

\author{Lucas Marcogliese\,\orcidlink{0009-0003-7097-0440}}
\author{Ouviyan Sabapathy}
\affiliation{
 JARA-FIT Institute for Quantum Information, Forschungszentrum J{\"u}lich GmbH and RWTH Aachen University, Aachen, Germany
 }
\author{Rudolf Richter}
\affiliation{
 Fakult{\"a}t für Physik, Universit{\"a}t Regensburg, 93040 Regensburg, Germany
}
\author{Jhih-Sian Tu}
\affiliation{
Helmholtz Nano Facility (HNF), Forschungszentrum J{\"u}lich, J{\"u}lich, Germany
}
\author{Dominique Bougeard}
\affiliation{
 Fakult{\"a}t für Physik, Universit{\"a}t Regensburg, 93040 Regensburg, Germany
}
\author{Lars R. Schreiber\,\orcidlink{0000-0003-0904-9612}}
 \email{lars.schreiber@physik.rwth-aachen.de}

\affiliation{
 JARA-FIT Institute for Quantum Information, Forschungszentrum J{\"u}lich GmbH and RWTH Aachen University, Aachen, Germany
 }
 \affiliation{ARQUE Systems GmbH, 52074 Aachen, Germany}

\begin{abstract}
Si/SiGe heterostructures on bulk Si substrates have been shown to host high fidelity electron spin qubits. Building a scalable quantum processor would, however, benefit from further improvement of critical material properties such as the valley splitting landscape. Flexible control of the strain field and the out-of-plane electric field $\mathcal{E}_z$ may be decisive for valley splitting enhancement in the presence of alloy disorder. We envision the Si/SiGe membrane as a versatile scientific platform for investigating intervalley scattering mechanisms which have thus far remained elusive in conventional Si/SiGe heterostructures and have the potential to yield favourable valley splitting distributions. Here, we report the fabrication of locally-etched, suspended SiGe/Si/SiGe membranes from two different heterostructures and apply the process to realize a spin qubit shuttling device on a membrane for future valley mapping experiments. The membranes have a thickness in the micrometer range and can be metallized to form a back-gate contact for extended control over the electric field. To probe their elastic properties, the membranes are stressed by loading with a profilometer stylus at room temperature. We distinguish between linear elastic and buckling modes, each offering new mechanisms through which strain can be coupled to spin qubits.

\end{abstract}

\flushbottom
\maketitle

\section{Introduction}\label{Introduction}

Electron spin qubits in Si/SiGe heterostructures satisfy many of the important requirements for building a quantum computer \cite{stano_review_2022, vandersypen_interfacing_2017} such as long coherence times \cite{struck_low-frequency_2020}, high gate fidelities \cite{yoneda_quantum-dot_2018,xue_quantum_2022, noiri_fast_2022, mills_two-qubit_2022},  compatibility with established industrial processes \cite{neyens_probing_2024, huckemann_industrially_2025, koch_industrial_2025}, potential for scalability \cite{kunne_spinbus_2024}, and operation at relatively high temperatures \cite{petit_universal_2020}. The realization of thousands of qubits necessary for quantum error correction \cite{hetenyi_tailoring_2024} calls for a high degree of material quality and homogeneity. The energy separation between the two low-lying valley states, known as valley splitting $E_\text{VS}$, is a critical material parameter for electron spin qubits in Si/SiGe. If $E_\text{VS}$ is locally too low, it compromises spin readout \cite{tagliaferri_impact_2018}, reduces dephasing times \cite{hollmann_large_2020}, and complicates qubit shuttling fidelity \cite{langrock_blueprint_2023, oda_suppressing_2024, volmer_mapping_2024, losert_strategies_2024, david_long_2024}. Several strategies to address low valley splitting have been proposed \cite{losert_practical_2023}, but are thus far limited either to control schemes for avoiding regions of low $E_\text{VS}$ \cite{losert_strategies_2024, nemeth_omnidirectional_2024, oda_suppressing_2024}, to growth techniques which increase the mean $E_\text{VS}$ but may not necessarily prevent low valley-splitting values \cite{mcjunkin_valley_2021,wuetz_atomic_2022, klos_atomistic_2024}, or by strict constraints on alloy disorder and heterostructure growing conditions \cite{mcjunkin_sige_2022}.

Conventional Si/SiGe heterostructures split the sixfold valley degeneracy in silicon into a fourfold and lower-energy twofold one through the tensile biaxial strain ($\varepsilon_{xx}$ and $\varepsilon_{yy}$) in the quantum well \cite{paul_sisige_2004}. The remaining twofold splitting between $z$-valleys can be engineered through the in-plane shear strain $\varepsilon_{xy}$, the out-of-plane electric field $\mathcal{E}_z$, and the confinement potential $U(z)$ \cite{thayil_theory_2025}. In conventional heterostructures, the dominant mechanism which determines $E_\text{VS}$ involves scattering between valleys within the same Brillouin zone (BZ). 

It has been proposed that intervalley scattering across different BZs will give rise to \textit{deterministic} enhancements which are less sensitive to atomistic disorder, that is yielding an $E_\text{VS}$ distribution with negligible weight near zero \cite{woods_coupling_2024, gradwohl_enhanced_2025}. Shear strain $\varepsilon_{xy}$, generated in $(001)$ Si/SiGe by applying uniaxial stress along the $[110]$ direction, is a key ingredient in accessing these inter-BZ scattering mechanisms. 

Strain has been practically introduced in various forms: deposited materials with tunable stress such as Si$_3$N$_4$ \cite{mackenzie_stress_2006, jaikissoon_cmos-compatible_2024}, selectively implanted regions \cite{sawano_origin_2013, masteghin_stress-strain_2021}, and piezoelectric materials \cite{kaleli_integration_2014,cheema_enhanced_2020}. Such strategies have already been shown to improve the mobility of MOSFETs and spin lifetime in Si spin field-effect transistors \cite{ sverdlov_silicon_2015, noborisaka_valley_2024, weber_experimental_2010}. In addition, Si/SiGe membranes enhance (shear) strain effects compared with bulk substrates \cite{scopece_straining_2014, schulli_dynamic_2024}.

In the context of spin qubits in Si/SiGe, a handful of  designs featuring shear strain have been proposed such as fin structures \cite{adelsberger_valley-free_2024} and stressor-membrane geometries involving silicon-on-insulator substrates \cite{woods_coupling_2024}. Membranes have the additional advantage of preserving a (locally) planar surface for the top-side device. 

These membranes also offer the possibility to control the out-of-plane electric field $\mathcal{E}_z$ independently of the quantum dot (QD) occupancy if the back-side is metallized \cite{ruggiero_backgate_2024}. $\mathcal{E}_z$ sets the envelope function's overlap with the Si/SiGe barrier, thereby increasing the valley splitting \cite{lima_interface_2023} involving both intra- and inter-BZ valley scattering \cite{thayil_theory_2025}, and allows for tuning spin-orbit coupling and associated intervalley effects \cite{nestoklon_electric_2008, prada_spinorbit_2011, woods_g-factor_2024}. 

In this work, we demonstrate the fabrication and discuss the mechanical properties of a Si/SiGe membrane serving as a future platform for studying and enhancing valley splitting with the electric field or shear strain. We identify two distinct elastic membrane modes that extend the strain engineering toolbox for Si/SiGe spin-qubit devices. Finally, we demonstrate process integration on a fully fabricated QuBus device \cite{struck_spin-epr-pair_2024, xue_sisige_2024} designed for valley mapping experiments \cite{volmer_mapping_2024, volmer_reduction_2025}.

%%%%%%%%%%%%%%%%%%%%%%%%%%%%%%%%%%%%%%%%%%%%%%%%%%%%%%%%%%%%%%%%%%%%%%%%%%%%%%%%%%%%%%%%%%%%%%%%%%%%%%%%

\section{Device model and design}\label{Design}

\begin{figure}[h]
    \centering
    \includegraphics[width=\linewidth]{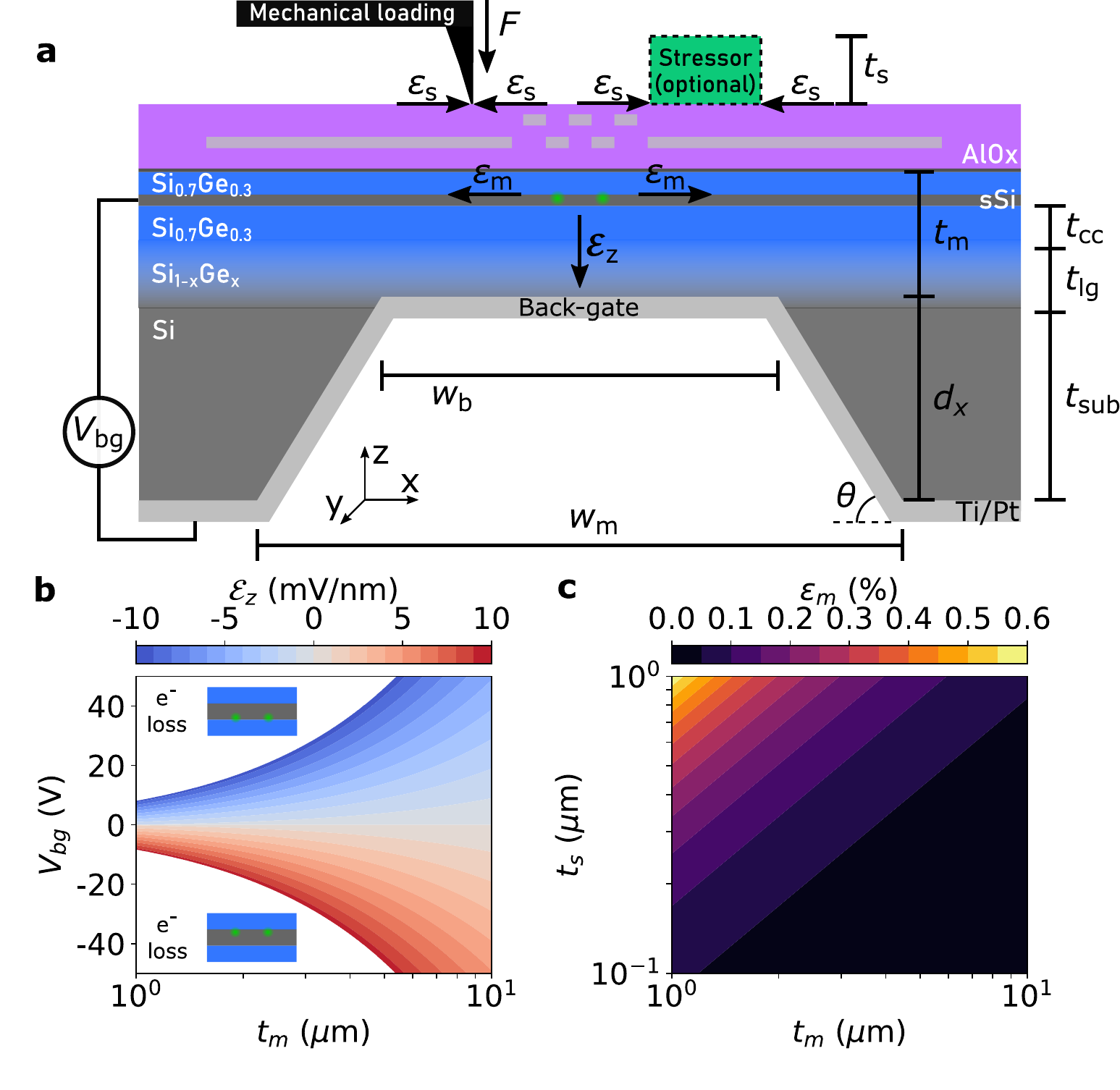}
    \caption{Device model and design. (a) Schematic (not to scale) of the geometry of the device with exemplary gate-defined quantum dots (QDs, green dots) in a suspended Si/SiGe membrane with metallic back gate and patterned multilayer front gates (bright gray). Strain may be coupled to the QDs via mechanical loading and/or a deposited stressor (e.g., Si$_3$N$_4$ grown by plasma-enhanced chemical vapor deposition) at the surface. The (virtual) substrate, consisting of handle wafer, linearly graded, and constant-composition buffers of thickness $t_{\text{sub}}, t_{\text{lg}}$ and $t_\text{cc}$, respectively, is locally etched at depth $d_x$. A square membrane of thickness $t_m$ is obtained with base width $w_b$ determined by mask width $w_m$. A mechanical load of force $F$, or stressor, with a thickness $t_s$, induces in-plane strain at the surface ($\varepsilon_s$) as well as in the quantum well ($\varepsilon_m$) hosted by the membrane. Note that $\varepsilon_m$ can be either compressive or tensile. “sSi” refers to (epitaxially) strained silicon. (b) Out-of-plane electric field $\mathcal{E}_z$ in the quantum well as a function of voltage $V_\text{bg}$ applied to the back gate. For negative (positive) $V_\text{bg}$, QDs form on the top (bottom) of the Si/SiGe quantum-well barriers. The white region corresponds to 99.9\% probability of electron escape from the QD via tunneling in the $z$ direction within one day. (c) Membrane strain $\varepsilon_m$ as function of stressor and membrane thicknesses.}
    \label{fig:1}
    \subfiglabel{fig:1a}{a}
    \subfiglabel{fig:1b}{b}
    \subfiglabel{fig:1c}{c}
\end{figure}

Here we outline the concept of our device, starting from a conventional Si/SiGe heterostructure. Typical Si/SiGe heterostructures, such as the one depicted schematically in Fig.~\ref{fig:1a} (outside the membrane region), are epitaxially grown on commercially available Si handle wafers with a thickness $t_\text{sub}$ two orders of magnitude larger than the Si/SiGe stack above it. A Si$_{1-x}$Ge$_x$ layer ($t_\text{lg}$) with linearly graded Ge concentration $x$ serves to transition from the Si lattice constant to that of SiGe with a constant composition ($t_\text{cc}$) without strain, forming a virtual SiGe substrate. In this way, the relatively thin Si layer hosting the QDs can be biaxially strained which is important for lowering the energy of the two $z$-valleys versus the four $x$-,$y$-valleys. Finally, a SiGe spacer layer completes the quantum well. 

The membrane device illustrated in Fig.~\ref{fig:1a} can be used as a versatile platform for enhancing the splitting between the remaining $z$-valleys via two possible paths: (1) electric field and (2) shear strain. The membrane can be metallized to form a back-gate contact that is close enough to the quantum well for extended control over $\mathcal{E}_z$ via the back-gate voltage $V_{\text{bg}}$ and appropriate top-side gate voltage offsets. It also offers the possibility for strain engineering either by a tunable mechanical load $F$ applied perpendicular to the surface or a fixed, rectangular stressor oriented along the $[110]$ direction with uniaxial strain $\varepsilon_s$. In either of these cases, the symmetry between in-plane 
strains can be broken, leading to a nonzero strain-tensor component $\varepsilon_{xy}\approx\varepsilon_m\neq0$, where $\varepsilon_m$ represents the in-plane strain component generated by the mechanical load or the deposited stressor at the depth of the quantum well.

To observe these effects, we must consider the necessary membrane geometry. The ideal etch process yields a square membrane with a base width $w_b$. As will be shown in the next section, $w_b$ can be made much larger than the quantum device to guarantee overlap. The sidewall angle $\theta$ is fixed by the etch process and has a negligible effect on both membrane strain and electric field seen by the QDs. Thus $w_b$ and $\theta$ are not critical for the foreseen device operation. Meanwhile, the membrane thickness $t_m$ depends on the substrate $t_\text{sub}$, linearly graded buffer $t_\text{lg}$ and constant-composition buffer $t_\text{cc}$ thicknesses as well as the etch depth $d_x$ as $ t_m \approx t_\text{sub} + t_\text{lg} + t_\text{cc} - d_x$, neglecting the much thinner quantum well and SiGe spacer. Note that $x$ refers to the local alloy fraction of Ge with respect to the etch front, which governs the evolution of the etch depth, as described in more detail in Section \ref{Fabrication}. The membrane thickness enhances both strain and the electric field nonlinearly. By adopting simple electrostatic and strain models, we determine the target membrane thickness $t_m$ for both these applications below.

First, we estimate the electric field $\mathcal{E}_z$ as a function of $t_m$ and back-gate voltage $V_\text{bg}$ as shown in Fig.~\ref{fig:1b}, assuming an offset for the voltages on the top-side gates (approximated as a single, continuous metal sheet) to maintain a constant QD occupancy. For reference, the electric field at $V_\text{bg}=0$\,V for typical top-side voltages on a conventional Si/SiGe heterostructure was simulated using \textsc{comsol multiphysics} to be on the order of 2\,mV/nm. In Fig.~\ref{fig:1b}, we assume that QDs can be formed on both the top and bottom edges of the Si quantum well, depending on the sign of $V_\text{bg}$. In either case, there is a maximum $\mathcal{E}_z$ (approximately 10\,mV/nm) that can be achieved beyond which the QD has a non-negligible probability of tunneling out of the quantum well (white region) in the $z$ direction. We estimate this probability according to the Wentzel-Kramers-Brillouin approximation \cite{fowler_electron_1928} with a target charge stability of 99.9\% for a full day. This allows us to define a lower bound on $t_m$ that depends on the maximum absolute gate voltage supported by the sample dielectrics and measurement setup. For typical heterostructures in the disorder-dominated regime \cite{losert_practical_2023}, both the mean and the variance of the valley splitting increase linearly with $\mathcal{E}_z$ at electric fields above about 4\,mV/nm \cite{boykin_valley_2004, lima_interface_2023}, thereby reducing the probability of hitting a region of low $E_\text{VS}$. Assuming we operate the back gate quasistatically and can apply up to $\pm$\,50\,V, we can reach the maximum $\mathcal{E}_z$ with a membrane thickness of $t_m\leq$ \SI{5}{\micro\meter}. 

Second, we determine the membrane thickness at which strain effects become relevant for valley splitting ($\varepsilon_{xy} \gtrsim 0.1$\,\%)  \cite{thayil_theory_2025}. For this purpose, we consider the case of a deposited Si$_3$N$_4$ stressor and estimate the strain sharing between the stressor and a Si$_{70}$Ge$_{30}$ membrane of thickness $t_s$ and $t_m$, respectively. We assume $t_m \ll w_b$ and that the membrane is composed entirely of unstrained Si$_{70}$Ge$_{30}$ for a conservative initial estimate of the required membrane thickness \cite{roberts_elastically_2006}. The strain $\varepsilon_m$ in the membrane plotted in Fig.~\ref{fig:1c} reaches $\varepsilon_{xy} \approx 0.1$\,\% at $t_\text{m} \approx $ \SI{3}{\micro\meter} for a Si$_3$N$_4$ stressor with $t_s=$ \SI{0.5}{\micro\meter} and residual compressive stress of 1 GPa which is readily achievable by plasma-enhanced chemical vapor deposition (PECVD). 

In the following section, we discuss the fabrication process used to realize a membrane meeting the above criteria.

\section{Fabrication}\label{Fabrication}

Membranes were fabricated from two heterostructures, hereafter referred to as wafer A and wafer B, grown by chemical vapor deposition and molecular beam epitaxy, respectively. Wafer A has a handle substrate thickness $t_\text{sub}$ of \SI{725}{\micro\meter} and an unpolished back side made smooth [less than 10\,nm root-mean-square (rms) roughness] by chemical-mechanical polishing, whereas heterostructure B was grown on a double-side polished 525-\SI{}{\micro\meter}-thick substrate. The heterostructures differ mainly in their virtual substrate with linearly graded Si$_{1-x}$Ge$_x$ ($t_\text{lg}$) and constant-composition Si$_{70}$Ge$_{30}$ ($t_{cc}$) buffer thicknesses $(t_\text{cc},t_\text{lg})$ of (2.3, 2.7)\,\SI{}{\micro\meter} and (0.5, 3.7)\,\SI{}{\micro\meter} for A and B, respectively. The rest of the layerstack above the virtual substrate has little influence on the membrane fabrication, but thicknesses are comparable between each wafer.  

The back sides of the substrates are coated with \SI{1}{\micro\meter} SiO$_2$ by PECVD serving as a hard mask for wet etching with tetramethylammonium hydroxide (TMAH). The hard mask is patterned with squares oriented along the $[110]$ direction on a $(001)$ surface using buffered oxide etch (BOE). Note that the $[110]$ direction corresponds to the common orientation used for qubit linear arrays and shuttling devices (e.g., Refs. \cite{xue_sisige_2024, struck_spin-epr-pair_2024}). The alignment of the hard mask with the given crystallographic direction yields an anisotropic etch profile due to the high selectivity to (111)-planes \cite{pal_high_2021} (see Appendix \ref{app:A}). For an ideal crystal, the sidewalls in Si have an angle of \SI{54.7}{\degree}, making it possible to metallize the back side with nonconformal techniques such as electron-beam evaporation to achieve electrical contact from the unetched surface down to the membrane for back-gate applications. 

We protect the top side of the samples during etching using a custom-made sample holder manufactured by Advanced Micromachining Tools GmbH which seals one surface away from the etchant using two rubber gaskets. The etching basin is heated to \SI{80}{\degreeCelsius} and covered by a lid to minimize changes in TMAH concentration (25\,\%). 

% The etch rates of Si and Si$_{70}$Ge$_{30}$ were measured on a Si reference sample and the topside of a SiGe heterostructure with $w_\text{m}=$ \SI{940}{\micro\meter} to be 441 and 5 nm/min respectively, consistent with previously reported values \cite{li_germanium_1999, loup_si_2013}. 

\begin{figure}[h]
    \centering
    \includegraphics[width=\linewidth]{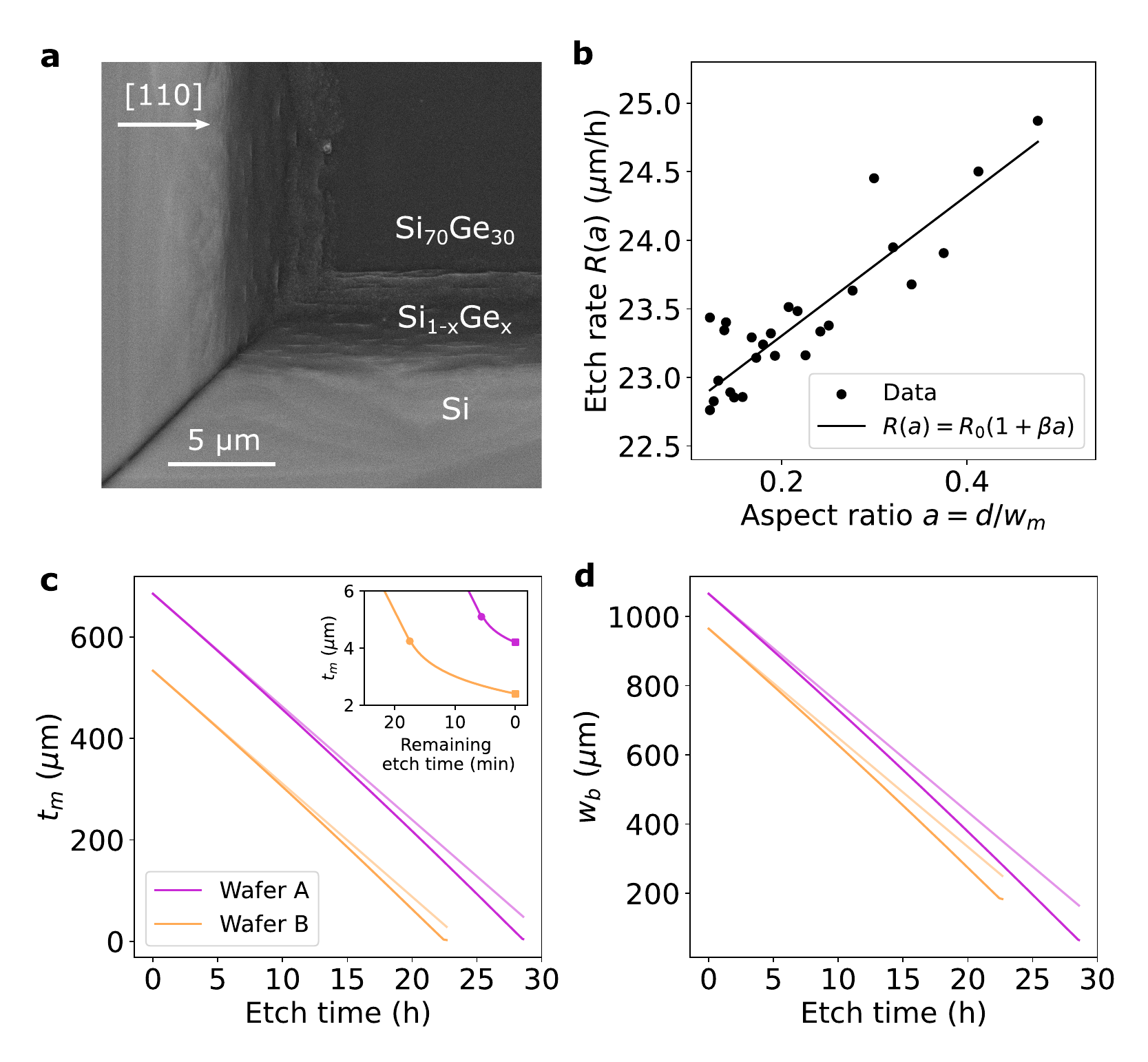}
    \caption{Aspect-ratio-dependent anisotropic etching. (a) Scanning electron microscope image of the membrane base corner, showing gradual change in the sidewall angle due to the linearly graded buffer. (b) Apparent etch rates $R(a)$ of silicon as a function of the aspect ratio $a=d_x/w_m$ for a fixed etch time. The solid line represents a least-squares fit of the empirical relation $R(a)=R_0(1+\beta a)$ where $R_0$ is the etch rate for $a=0$ and $\beta$ is a dimensionless factor which determines the strength of aspect-ratio-dependent effects. (c) Membrane thickness $t_m$ and (d) base width $w_b$ as a function of time as predicted by the etch-rate model. Faded lines depict a linear-etch-rate assumption $d_x(t) \approx R_0t$. The inset of (c) shows $t_m$ within 25 minutes of the total etch times for wafers A and B (squares), with the start of the linearly graded buffer indicated by dots.}
    \label{fig:2}
    \subfiglabel{fig:2a}{a}
    \subfiglabel{fig:2b}{b}
    \subfiglabel{fig:2c}{c}
    \subfiglabel{fig:2d}{d}
\end{figure}

The accuracy and reproducibility of the fabrication process rely, to first order, on the high selectivity between Si and Si$_{70}$Ge$_{30}$ \cite{li_germanium_1999, loup_si_2013}. The difference in etch rate of nearly two orders of magnitude implies that the SiGe virtual substrate acts as an effective etch stop. We note that the sidewalls in Si$_{70}$Ge$_{30}$ appear to have an angle of approximately \SI{35}{\degree}, well in accordance with the theoretical angle between (001)- and (211)-planes. As a result, the sidewall angle is expected to shift monotonically from \SI{54.7}{\degree} to \SI{35}{\degree} within the linearly graded buffer [Fig.~\ref{fig:2a}], which is important to avoid shadowing effects during metallization. 

Despite the selectivity between Si and Si$_{70}$Ge$_{30}$, assuming a constant etch rate would lead to incorrect and unreliable results. Analogously to many reactive ion etching processes \cite{wu_high_2010, rangelow_critical_2003}, we observe that wet etching with TMAH is subject to aspect-ratio-dependent effects (ARDE). The term “ARDE” describes the change in etch rate as a function of aspect ratio resulting in an etch rate that is nonlinear in time \cite{yeom_maximum_2005, lai_aspect_2006}.

To characterize ARDE in TMAH-etched silicon and develop an etch-rate model for precise geometry control, we conduct two tests. The first involves a sample with an array of mask widths $w_m$ on 525-\SI{}{\micro\meter}-thick silicon substrates etched for a fixed time $\Delta T$ of 20 hours 55 minutes. According to Fig.~\ref{fig:2b}, the apparent etch rate $R(a)=\frac{d_x}{\Delta T}$ increases with the aspect ratio $a=\frac{d_x}{w_m}$. The second is a sample with a $3\times3$ array of fixed $w_m$ and the same etch time, in which we observed a very small deviation of etch depth despite large variations in membrane base width $w_b$ (nonideal relation between $w_m$ and $w_b$ discussed further in Section \ref{Fabrication}). From these observations, we can draw the following conclusions: (1) the variation of the etch rate with depth cannot depend directly on $w_b$; (2) it is difficult to imagine a physical mechanism which explains a direct dependence of the etch rate on $w_m$ other than through the aspect ratio $a$. These trends point to a diffusion-limited process in which the TMAH concentration is locally depleted close to the membrane surface. Since the etch rate increases with decreasing TMAH concentration \cite{loup_si_2013}, the etch rate accelerates as the etch front proceeds deeper into the substrate.

By assuming that the ARDE are stationary ($dR(a)/dt$ is constant), we can equate the empirical linear trend in Fig.~\ref{fig:2b} with the instantaneous etch rate:

\begin{align}
    R(a(t)) = \frac{d}{dt}[w_m a(t)] = R_0(1+\beta a(t)),
\end{align}

\noindent where $R_0$ is the etch rate at $a=0$ and $\beta$ is a dimensionless rate at which the etch rate changes with $a$. This differential equation can be solved to obtain the etch depth as a function of time in silicon. However, in Si/SiGe, we must also take the varying Ge concentration $x$ as a function depth into account. We assume an exponential dependence of $R_0$ with $x$ \cite{loup_si_2013} such that the general expression for instantaneous etch rate becomes:

\begin{align}
    R(x(t),d_x(t)) = \frac{d}{dt}[d_x(t)]= R_0e^{\gamma d_x(t)} \left(1+ \frac{\beta}{w_m}d_x(t)\right),
\end{align}

\noindent where $\gamma = \frac{1}{t_\text{lg}}\ln\left({\frac{R_{30}}{R_0}}\right)$ depends on the zero-aspect-ratio etch rates $R_x$ of Si ($R_{0}$) and Si$_{70}$Ge$_{30}$ ($R_{30}$). The base width is then simply given by $w_b(t) = w_m - 2d_x(t)\cot(\theta_x)$ with $\theta_x$ also scaling exponentially with $x$ in the linearly graded buffer. In Fig.~\ref{fig:2c}, the evolution of the membrane thickness $t_m(t) = t_\text{sub} + t_\text{lg} + t_\text{cc} - d_x(t)$ according to the model (dark lines) for both wafers is compared with $t_m$ obtained by assuming a constant etch rate $R(a(t))=R_0$ (faded lines). We emphasize that assuming a constant etch rate greatly increases the risk of etching through the entire heterostructure because the overetch time that results from underestimating the etch rate is comparable to the time it takes to etch through the rest of the SiGe.

Fig.~\ref{fig:2d} illustrates the evolution of the base width $w_b$ whose dependence on $d_x(t)$ also causes the ARDE etch-rate model (solid lines) to deviate from the constant-etch-rate assumption (faded lines). By carefully choosing $w_m$, we obtain excellent control of the etched base width $w_b$ and the sidewall angle $\theta$ on bare silicon samples (see Appendix \ref{app:A}). In the next section, we apply the model to target specific membrane geometries $(t_m, w_b)$ on the two wafers.

In summary, our control of the etch rate enables us to target membrane thicknesses $t_m$ with an accuracy in the micrometer range meeting the design requirements. With the determined etch rates, we etched a 10$\times$10\,mm$^2$ sample from each wafer with 3$\times$3 arrays of square masks with target widths $w_b$ of 200 and \SI{300}{\micro\meter} for 30 h 45 min and 22 h 40 min, respectively. These etch times were chosen with the objective of reaching the linearly graded buffer without etching through it completely [as shown in Fig.~\ref{fig:2c}]. With the different target aspect ratios $t_m/w_b$, we intended to match the typical dimensions of the etched mesa in a shuttling device and to push the limits of the lateral dimension of membranes in order to accommodate larger strain. We then characterized the resulting Si/SiGe membranes with the methods presented below.

\section{Characterization}\label{Characterization}

Next we elaborate on the methods used to characterize the geometry of the membranes and validate their compatibility with the fabrication of spin-qubit devices. 

We used an Accurion EP4 ellipsometer with a lateral resolution of \SI{1}{\micro\meter} to measure the membrane thickness from the top side, averaging over approximately $50$\,\% of the membrane area. The oscillations observed in both the amplitude and phase of the complex reflectance ratio in the red part of the visible spectrum are a signature of thin-film interference indicating that the membrane thicknesses are comparable to the wavelength of the incident light (see Appendix \ref{app:B}). The transfer matrix method provides a convenient formalism to treat multilayer structures containing graded index layers such as the SiGe buffer. The \textsc{tmm} and \textsc{tmm-fast} Python packages are used to generate the ellipsometric models \cite{byrnes_multilayer_2020, luce_2022}. Since the total thickness of the membrane is comparable to the penetration depth of the light, air is defined as the incoming and outgoing semi-infinite medium. 

\begin{figure}[h]
    \centering
    \includegraphics[trim={0.8cm 0.7cm 0 0},clip,width=.95\linewidth]{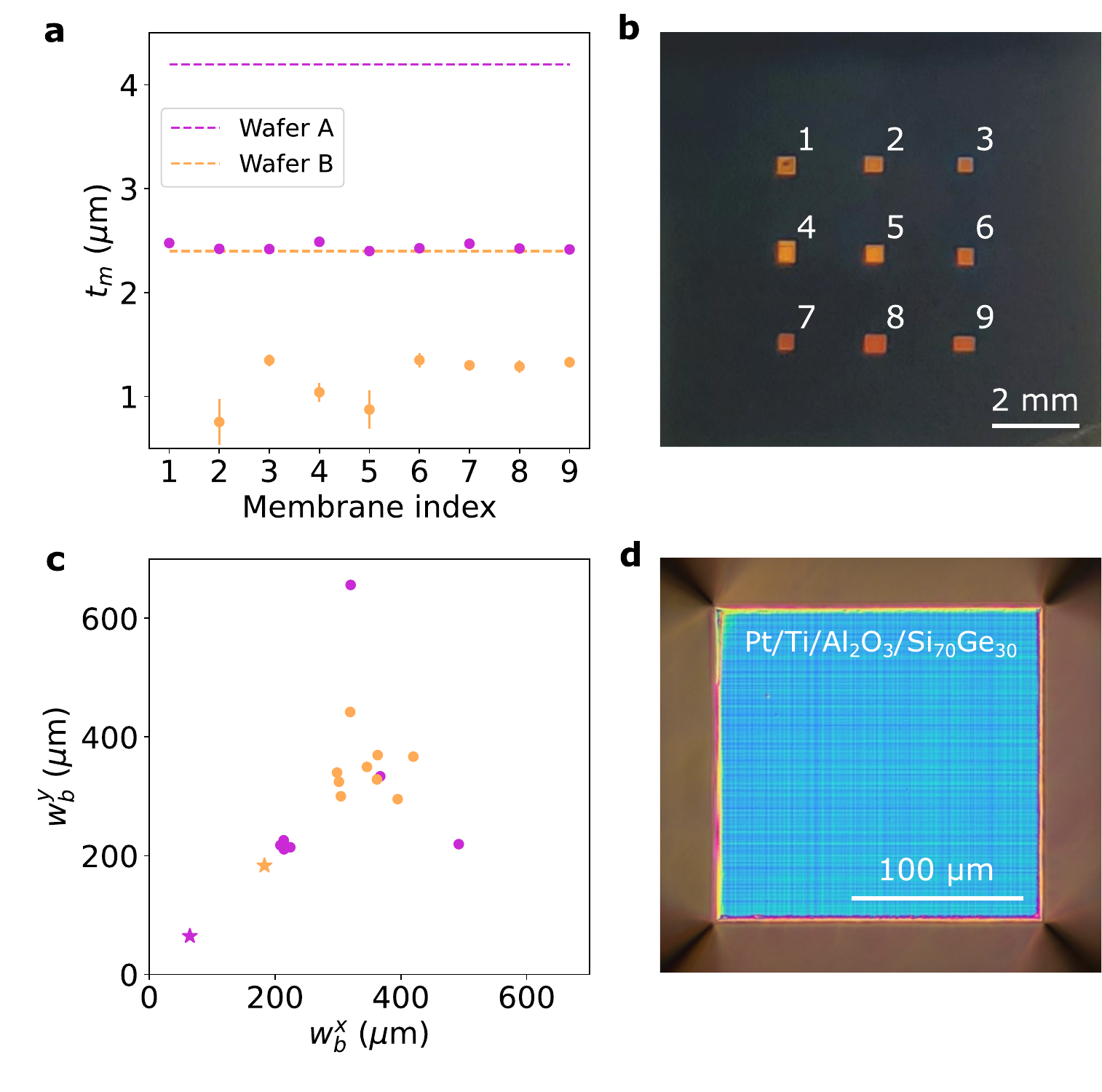}
    \caption{Observed membrane geometry. (a) Membrane thickness $t_m$ measured by spectroscopic ellipsometry (dots) and target thickness calculated by the etch-rate model (dashed lines). (b) Optical microscope image of a 10$\times$10 mm$^2$ sample from wafer B containing a 3$\times$3 array of partially transparent membranes.  (c) Membrane widths $w_b$ along $x$ and $y$ obtained by optical microscopy for wafers A and B. Targeted widths obtained from the etch-rate model are depicted by stars. (d) Optical polarized microscope image of etched substrate from the back side with visible cross-hatch pattern. }
    \label{fig:3}
    \subfiglabel{fig:3a}{a}
    \subfiglabel{fig:3b}{b}
    \subfiglabel{fig:3c}{c}
    \subfiglabel{fig:3d}{d}
\end{figure}

The resulting fits in Fig.~\ref{fig:3a} yield thicknesses of 2.44(1) and 1.25(5)\,$\SI{}{\micro\meter}$ for wafers A and B respectively within reasonable agreement with the etch-rate model (dashed lines). The uncertainty in the etch-rate model is likely attributable to the indirect estimation of the etch rate in the linearly graded buffer as well as the possible influence of the threading dislocations in the buffer on the etch rate. The periodicity in the measured thicknesses of membranes from wafer A is presumably related to the vertical orientation of the sample in the etch basin. The hydrogen gas generated as a byproduct of the etching process can accumulate more on higher surfaces of the sample, slightly lowering the etch rate of the upper rows of the $3\times 3$ membrane arrays as a result. The ellipsometric data from membrane 1 on wafer B lacks oscillations likely due to features in topography explained in Section \ref{Load deformation}. Membranes 2, 4, and 5 also appear thinner and, together with membrane 1, form the upper left side of the sample which might have experienced local variations during etching. 

\begin{figure}[h]
    \centering
    \includegraphics[trim={0 0 0.7cm 0},clip,width=\linewidth]{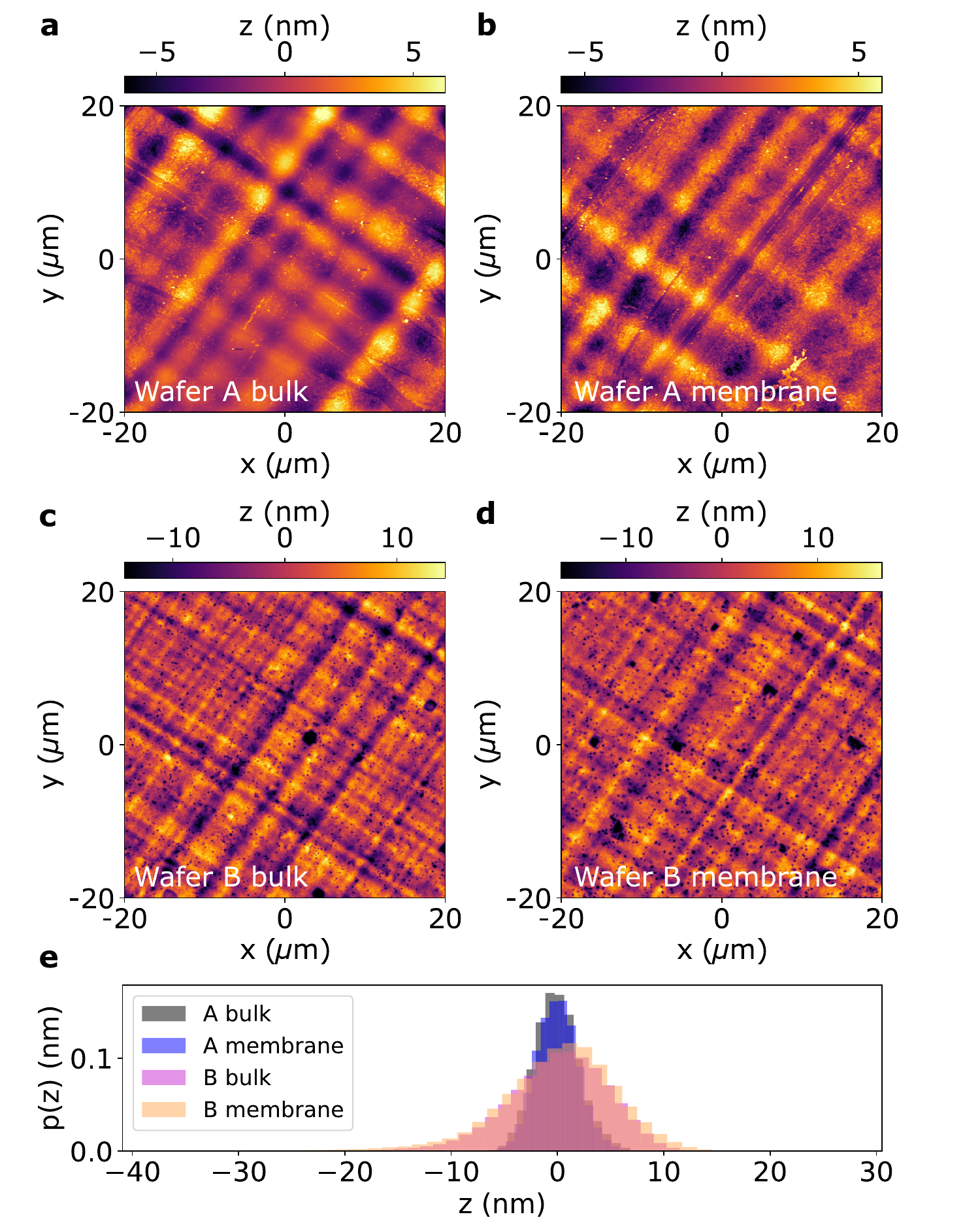}
    \caption{Surface roughness by atomic force microscopy. (a),(b) Surface topography of wafer A for the bulk (a) and membrane (b), region, respectively. Background curvature is subtracted by a sixth-order polynomial and horizontal scars are corrected by the open-source software \textsc{gwyddion} \cite{necas_gwyddion_2012}. (c),(d) Same for wafer B with the bulk (c) and membrane (d) region, respectively. (e) Probability density of the measured height profiles from (a)-(d).}
    \label{fig:4}
    \subfiglabel{fig:4a}{a}
    \subfiglabel{fig:4b}{b}
    \subfiglabel{fig:4c}{c}
    \subfiglabel{fig:4d}{d}
    \subfiglabel{fig:4e}{e}
\end{figure}

\begin{figure*}[!t]
    \centering
    \includegraphics[width=\linewidth]{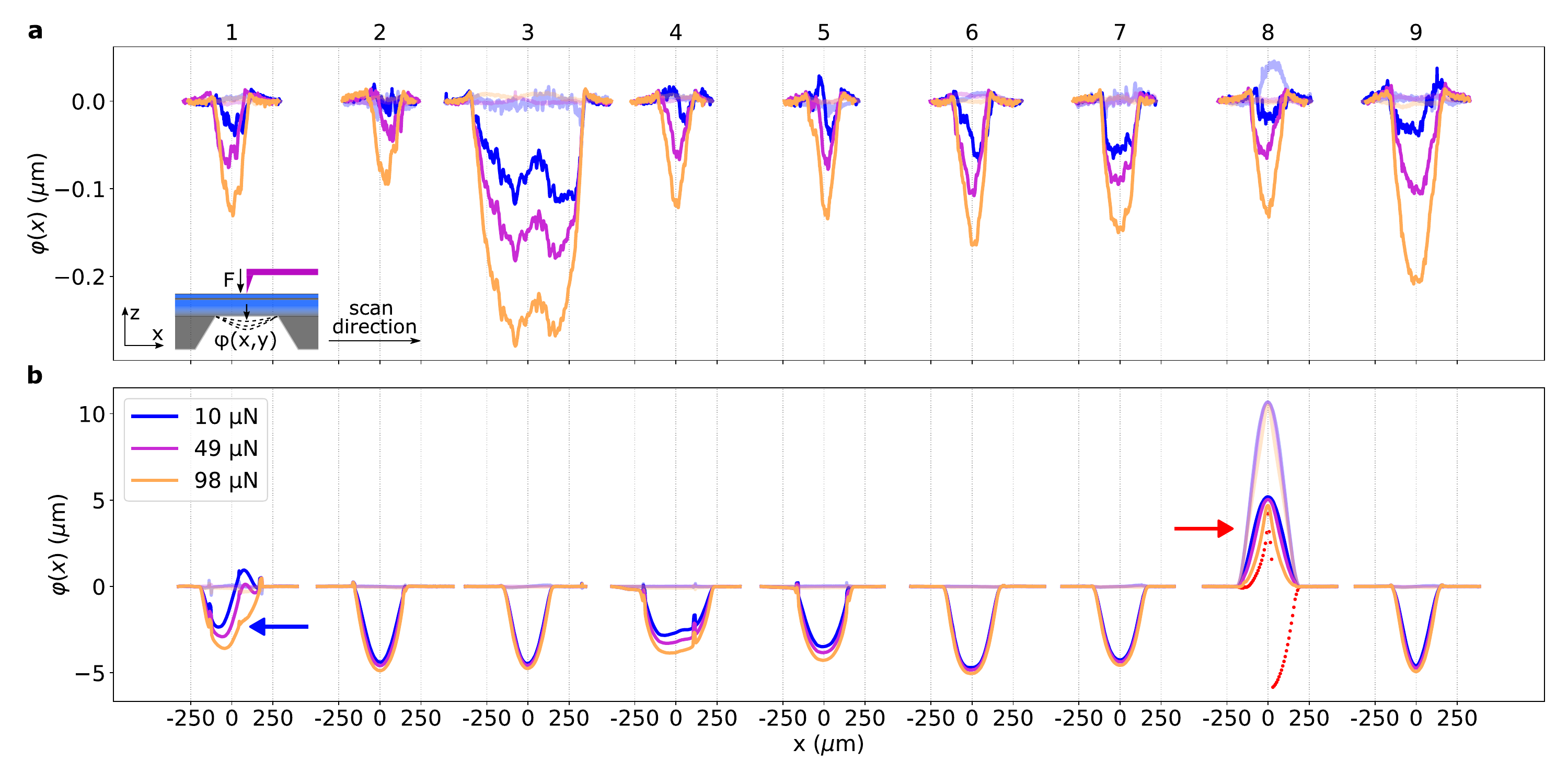}
    \caption{Membrane elasticity. (a),(b) Membrane vertical deformations $\varphi(x)$ measured by profilometry with increasing force $F_\nearrow = \{9.8, 49, 98\}$ \SI{}{\micro\newton} corresponding to blue, pink and yellow curves respectively for wafers A (a) and B (b). Each set of curves corresponds to a different membrane on the sample, offset horizontally for better readability. Differences in membrane profiles for increasing $F_\nearrow$ and decreasing $F_\searrow$ scans are shown in partially transparent curves with corresponding colors. }
    \label{fig:5}
    \subfiglabel{fig:5a}{a}
    \subfiglabel{fig:5b}{b}
\end{figure*}

Given that silicon has an absorption depth ranging from about 100\,nm to \SI{10}{\micro\meter} within the visible spectrum \cite{green_optical_1995}, the membranes are partially transparent, even to the naked eye [Fig.~\ref{fig:3b}]. This allows us to measure the membrane widths $w_\text{b}$ along $x$ and $y$ by an optical microscope from the top side. The spread in $w_b$ across the $3\times3$ array of membranes within a sample and the deviation from the model-calculated $w_b$ (star data points) are noticeably broader, particularly for wafer B [Fig.~\ref{fig:3c}] than for the silicon reference wafer (Appendix \ref{app:A}). We speculate these discrepancies to be a consequence of a larger dicing misalignment of the heterostructure wafers compared with the silicon reference sample. The self-alignment of the etch front with the etch-selective planes appears to considerably increase the effective initial mask width $w_m$ (Appendix \ref{app:A}). Meanwhile, the measured thicknesses are closer to predicted values because the etch rate depends only on the physical hard mask width $w_m$ which behaves like an aperture with respect to etchant diffusion.

We metallized the back side of the membranes with 30 nm Al$_2$O$_3$, 5 nm Ti and 150 nm Pt by electron-beam evaporation after removal of the hard mask in BOE. Fig.~\ref{fig:3d} shows a polarized microscope image of one such membrane from wafer B. Using a dc probe station, we confirmed electrical contact between all combinations of unetched, sidewall, and membrane surfaces. Remarkably, the membranes retain their structural integrity after contact with the probe needles (uncontrolled applied pressure), and even survive the mechanically demanding dicing process used to singulate devices for measurement. 

Fig.~\ref{fig:3d} also reveals a cross-hatch pattern visible from the back side of a membrane after metallization and provides confirmation that the substrate has been etched into the linearly graded buffer. We also checked the top side of the membranes for changes to surface morphology as a result of etching away the bulk substrate beneath the membrane. For reference, we compare the surface of the membranes and the bulk of both wafers by atomic force microscopy. The atomic force microscopy maps of bulk and membranes of wafer A [\figsubref{fig:4}{fig:4a,fig:4b}] and B [\figsubref{fig:4}{fig:4c,fig:4d}] and the corresponding surface height distributions [Fig.~\ref{fig:4e}] show little change on either wafer between bulk and membrane. The deeper dots in wafer B are presumably intersecting dislocation networks at the silicon initiation buffer. 

The rms roughness of the surface of wafer A remains nearly the same (2.1\,nm for bulk and 2.0\,nm for membrane), while wafer B sees a slight increase from 4.8 to 5.6\,nm after etching, possibly due to buckling of the membrane (see Section \ref{Load deformation}). We also measured the rms roughness of the back side of the membranes to be below the 8\,nm resolution of the profilometry scan range required to reach the bottom of the etched structure. Thus, the surface topography is suitable for spin-qubit device fabrication as well as the implementation of a back-gate contact.

\section{Load deformation}\label{Load deformation}

To study the surface properties further and characterize the load response of the membranes, the topography $\varphi(x,y)$ of the samples was measured with a Bruker DektakXT profilometer. Measurements were carried out first with increasing ($F_\nearrow$) stylus load from 9.8 to \SI{147}{\micro\newton}, followed by applying forces in reverse order ($F_\searrow$) across the top side of the membranes. Exemplary $F_\searrow$ load deformations for $F_\nearrow \,= \{9.8, 49, 98\}$ \SI{}{\micro\newton} from both wafers are plotted in Fig.~\ref{fig:5}. 

We observe that the difference between $F_\nearrow$ and $F_\searrow$ scans, represented by faded lines, is negligible for all but one membrane from wafer B whose concavity changed during the measurement of membrane 8 with a force of \SI{147}{\micro\newton} illustrated by the red data points. This is an example of a buckling mode described in more detail below. The reversibility of the deflection and the absence of plastic deformation demonstrate that the membranes have an elastic load response and suggest that the structures are still comfortably away from the rupture point, thereby satisfying an important criterion for flexible strain engineering in these structures. 

\begin{figure}[h]
    \centering
    \includegraphics[width=\linewidth]{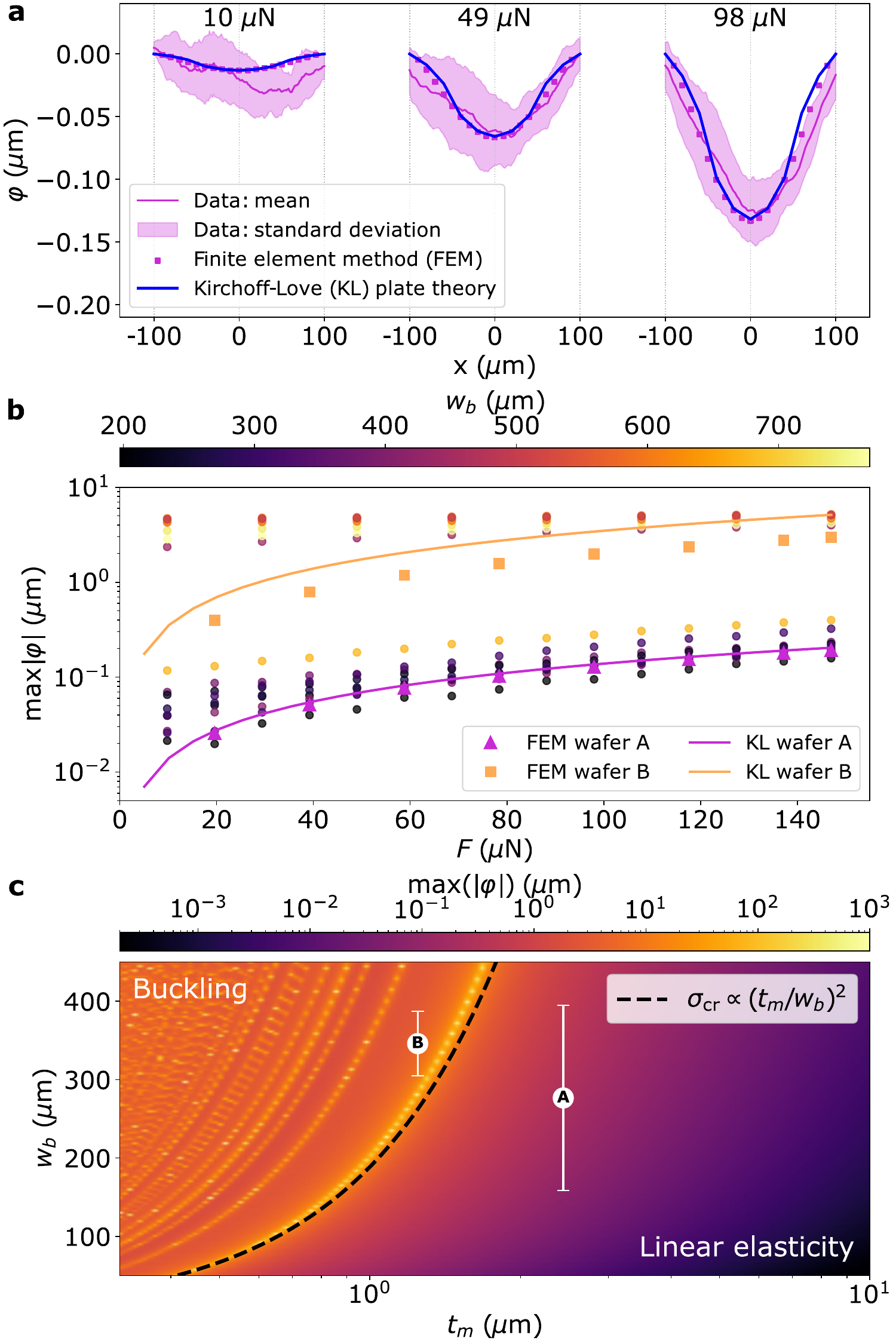}
    \caption{Simulation of deformation profiles. (a) Profilometry measurement, simulation by finite-element method (FEM, pink dots) and model by Kirchhoff-Love (KL) plate theory (blue solid lines) of the membrane deformation profile for different load positions across a centered linecut on wafer A. Standard deviation of the measured profiles shaded in pink. (b) Maximum vertical displacement, $\max{|\varphi|}$, of the profilometer stylus for wafers A (filled dots) and B (open dots) as function of width and mechanical loading. The FEM simulation (triangles and squares) and KL model (solid lines) assume linear elasticity of the membranes. (c) Color-coded maximum of the absolute value of the deformation profile, $\max{|\varphi|}$, as a function of membrane geometric parameters $t_m$ and $w_b$ calculated by KL plate theory. The dotted line represents the first buckling mode and traces the boundary between the linear elastic and buckling regimes.}
    \label{fig:6}
    \subfiglabel{fig:6a}{a}
    \subfiglabel{fig:6b}{b}
    \subfiglabel{fig:6c}{c}
\end{figure}

To distinguish between linear and nonlinear elasticity, we performed finite-element modeling (FEM) simulations using the Structural Mechanics module in \textsc{comsol multiphysics}. Convergence tests were conducted to reduce model volume to an effective geometry with a surface area of 1$\times$1\,mm$^2$ and a handle wafer thickness of \SI{62.54}{\micro\meter}. Vegard's law was applied to interpolate the elastic constants \cite{levinstejn_properties_2001} of the Si$_{1-x}$Ge$_x$ linearly graded buffer and the temperature was assumed to be constant at 300\,K.

The simulated profilometer tip positions $\varphi(x)$ in Fig.~\ref{fig:6a} agree well for membranes of wafer A, deviating only slightly for low applied forces. We note that the asymmetry in the data is presumably due to the scan direction and the convolution of the tip shape with the sample surface which are not captured by the FEM simulation. The simulated maximum absolute vertical displacement, $\max{|\varphi|}$, is indeed linear and Fig.~\ref{fig:6b} shows that  the data follows the same trend down to at least \SI{40}{\micro\newton} for membranes with widths close to the simulated $w_b=$ \SI{200}{\micro\meter}. Membrane A can therefore be modeled without requiring postbuckling analysis which is practical for device design and future strain engineering, including passive deposited stressors and active tunable loads.

Furthermore, as can be seen in Fig.~\ref{fig:5a}, the rms roughness (on top of background curvature) for weak loads is considerably larger than that measured by atomic force microscopy because the latter is limited to correlation lengths within the scan range of \SI{40}{\micro\meter}. We speculate that this long-range roughness could be a manifestation of \textit{local} buckling. The removal of the bulk substrate could in principle allow dislocation-induced strain to relax which is known to be correlated with the cross-hatch roughness \cite{degli_esposti_low_2024}. Local changes in topography could be one possible channel for such relaxation to occur.

In contrast, we observe much larger deformations in wafer B even for small applied mechanical loads. In some instances [Fig.~\ref{fig:5b}], the membranes in wafer B have concavity of opposite sign (red arrow, membrane 8) or even contain higher harmonics (blue arrow, membrane 1). The maximum absolute vertical displacement, $\max{|\varphi|}$, in Fig.~\ref{fig:6b} clearly does not match the FEM simulations which assume a linear elastic material. The responsible effect, called buckling, describes a sudden reduction in the stiffness of the structure leading to large, nonlinear deformations. This mechanism is triggered when in-plane compressive forces exceed the bending restoring force of the membrane \cite{becque_linking_2021}. To understand the origin of these compressive forces, let us first consider the case of a Si/SiGe heterostructure before back-side etching. The linearly graded and constant-composition SiGe buffers are nominally relaxed and essentially unperturbed by the comparatively thin, epitaxially strained Si quantum well. After back-side etching, the free surface beneath the SiGe buffer allows for a reequilibration of forces since the SiGe is no longer bound by an epitaxial relationship to the thick silicon substrate. The result is a compressively stressed SiGe buffer and spacer caused by strain sharing similar to that described in Ref. \cite{roberts_elastically_2006}. Given that the membrane is mostly made up of SiGe, its deformation and buckling threshold will be largely dictated by the material properties of SiGe. We can thus approximate the membrane as being composed entirely of SiGe and incorporate the effect of the strained quantum well as an initial compressive biaxial stress $\sigma_{\text{biax}}=\sigma_{xx}=\sigma_{yy}$. When these compressive forces, which in turn depend on the geometry of the membrane $(t_m, w_b)$ through strain sharing with the quantum well, exceed a critical limit granted by the membrane bending restoring force, the membrane will buckle across its entire thickness (in contrast to the surface effect of local buckling). 

The onset of global buckling is governed by the critical in-plane stress given by \cite{becque_linking_2021}:

\begin{align}\label{eq:buckling_cond}
    \sigma_{\text{cr}} = 2\widetilde{D}_{\text{ani}}\left(\frac{\pi t_m }{w_b} \right)^2, 
\end{align}

\noindent where $\widetilde{D}_{\text{ani}}$ is the generalized anisotropic flexural rigidity which depends on the material properties and crystal symmetry as elucidated in Ref. \cite{thomsen_silicon_2014}. Fig.~\ref{fig:6c} shows the different mechanical phases of a membrane as a function of geometric parameters $w_b$ and $t_m$ with an applied load of \SI{98}{\micro\newton} using Kirchhoff-Love (KL) plate theory (see Appendix \ref{app:C}). We assume the strain in the quantum well to be the same for both wafers A and B. For small enough $w_b$ and large enough $t_m$ as is the case for wafer A, the compressive stress in the SiGe membrane is below the critical threshold. It thus does not (globally) buckle and behaves instead as a linear elastic material, with deflections following a $t_m^3/w_b^2$ dependence consistent with linear FEM simulations. Larger and/or thinner membranes such as wafer B incur more compressive stress from the quantum well and exhibit \textit{global} buckling. As a result, the membrane deformations no longer obey linear elastic theory and feature one or more buckling nodes (undulation periods along a given in-plane direction). We note that since KL theory does not account for out-of-plane shear stresses $\varepsilon_{xz},\, \varepsilon_{yz}$, the KL deformations in Fig.~\ref{fig:6} are only an approximation. The linear elastic approximation holds for small applied loads (wafer A) but can only predict the shape of the buckling modes in the nonlinear regime, not the amplitude of their deflections. We also emphasize that these results are not a direct consequence of differences between wafers A and B but rather related to the distinct geometries of the membranes.

Whether linear or nonlinear (buckled) elastic regimes are preferable in the context of strain engineering for deterministic valley-splitting enhancement strategies remains to be verified experimentally. On the one hand, buckled membranes intrinsically involve more strain to support the larger displacements in the absence of external loading and can withstand greater loading without plastic deformation compared with linear elastic membranes. On the other hand, significant bowing could pose a risk to top-side structures such as nanoscale gates as well as complicate alignment between fine-gate layers.

\section{Quantum device integration}\label{Device_integration}

\begin{figure}[h]
    \centering
    \includegraphics[width=\linewidth]{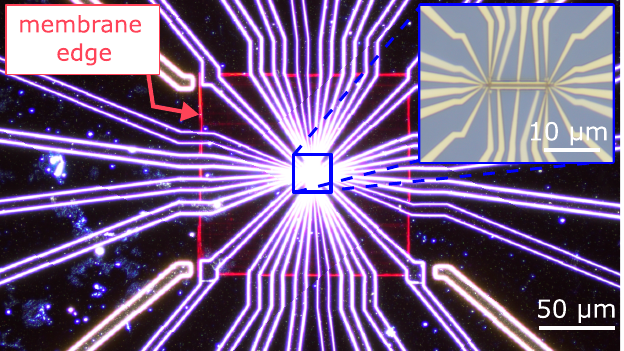}
    \caption{Fully fabricated spin-qubit shuttle on a back-gated membrane. Top-side dark-field image of a fully fabricated shuttling device through which we can see the etched membrane edges (red square). The shape of the metallic electrodes becomes visible by light reflected at their sidewalls. Inset: bright-field image of the spin-qubit shuttle device within the blue square of the dark-field image.}
    \label{fig:7}
\end{figure}

Having developed a fabrication process which can achieve Si/SiGe membranes with specific geometries and mechanical properties, we demonstrate the integration of the membrane with a QuBus device designed for shuttling-based valley mapping experiments \cite{volmer_mapping_2024, volmer_reduction_2025}. We fabricate the device on a substrate nominally identical to and with the same target membrane geometry as wafer A. The added complexity due to this process integration arises from: (1) hard mask deposition and lithography; (2) back-side alignment (BSA); (3) top-side surface/device protection; (4) relative membrane fragility compared with bulk substrates.

Besides hard mask deposition, all membrane-related processing is performed after top-side device fabrication to minimize the number of steps on the relatively fragile membrane. The top side is protected with resist before starting back-side hard mask lithography and remains on the sample during TMAH etching, hard mask removal, back-gate metallization, and dicing. In addition, we use a custom sample holder to seal away the top-side surface during TMAH etching since TMAH readily attacks resist. 

BSA is achieved with a Heidelberg Instruments DWL 66$+$ laser lithography tool with measured writing and calibration misalignments $(\Delta x, \Delta y)$ of (-0.8, 1.6)\,\SI{}{\micro\meter} and (2.0, -0.2)\,\SI{}{\micro\meter} respectively. However, the membrane alignment is ultimately dominated by the offset between the hard mask edge and the $\left[110 \right]$ direction. During etching, the etch front self-aligns with the $\left[110 \right]$ direction through an uncontrolled and often asymmetric rotation yielding rectangular etch bases that are marginally off-center. In the future, the process can be improved by first etching alignment markers with TMAH on the back side to which the rest of the lithography markers are aligned \cite{finnegan_high_2019}. However, in spite of the misalignment, the membrane can be made large enough to guarantee full overlap with the active area of typical quantum chips (around \SI{10}{\micro\meter}) and satisfy design requirements.

Using dark-field optical microscopy (Fig.~\ref{fig:7}), it is possible to see a membrane and its alignment with a fully fabricated shuttling device from the top side of a sample from wafer A. The inset confirms that the top-side gates remain intact after back-side processing, validating successful integration of the membrane process.

\section{Conclusion}\label{Conclusion}

With the objective of ultimately studying the effect of electric fields and strain on Si/SiGe spin-based quantum devices, we demonstrate the fabrication of a Si$_{70}$Ge$_{30}$/Si/Si$_{70}$Ge$_{30}$ membrane with an area of more than $100\times100$\,\SI{}{\micro\meter}$^2$ hosting a tensile strained Si quantum well. A proximal metallic back electrode and structured multilayer top electrodes are added to enable the formation of multiple quantum dots for which the electron occupation is independent of the out-of-plane electric field. We have optimized the fabrication steps, gate alignment, and reproducibility of the device fabrication. Membranes are characterized in terms of their geometry, surface roughness, and cross-hatch pattern typical of Si/SiGe heterostructures. The thickness of the membranes determines the elastic properties and response to mechanical loading procedures, which are both well captured by our finite-element modeling simulations and plate theory model. With membrane
thicknesses of 2.44(1) and 1.25(5)\,\SI{}{\micro\meter}, we obtain membranes in a linear or nonlinear elastic response regime offering us different pathways to strain devices with much more flexibility than on bulk substrates. The platform can be complemented by optional stressor materials. This new platform will allow us to investigate the dependence of the mapped local valley splitting on electric fields and strain at 100\,mK operation temperature \cite{xue_sisige_2024, volmer_mapping_2024, volmer_reduction_2025}. The valley splitting does not require individual fine tuning for each qubit. It is sufficient to achieve an increase in the minimum value across the qubit array. Since the membrane or back gate would be much larger than the individual qubits, they would have a global effect on the valley splitting of the qubits. It is, in principle, possible to host an entire two-dimensional shuttling array on a single membrane, provided that the fanout can route bond pads onto a bulk surface. Meanwhile, stressors can be patterned on the membrane surface to introduce asymmetries along the qubit array. By applying the valley mapping technique, we expect to be able to resolve these gradients to improve our knowledge of valley physics in the presence of strain. Understanding and achieving sufficiently large, local valley splitting is one of the most pressing material challenges to realize a scalable quantum computer in Si/SiGe.

\begin{acknowledgments}

The device fabrication was carried out at Helmholtz Nano Facility (HNF), Research Center J{\"u}lich GmbH \cite{albrecht_hnf_2017}. We thank the HNF staff, including Stefan Trellenkamp and Florian Lentz for e-beam lithography, Thomas Grap for chemical-mechanical polishing, Anja Zass and Kevin Rosenberger for ellipsometry measurements, Franz Josef Hackem{\"u}ller for support with back-side alignment, Luisa Koke and Lars Petter for hard mask depositions, Natalie Bruger and Stephany Bunte for dicing, the HNF evaporation team for metallization and Georg Mathey and Niklas Dörenkamp for their guidance regarding wet etching. We also thank Advanced Micromachining Tools GmbH for designing the custom sample holder. This work has been funded by the Deutsche Forschungsgemeinschaft (DFG, German Research Foundation) via Project-ID No. 289786932 (BO 3140/4-2 and SCHR 1404/2-2).

\end{acknowledgments}

\section*{Data Availability}
The data that support the findings of this study are openly available on Zenodo \cite{marcogliese_sige_membranes_2025}. 

%%%%%%%%%%%%%%%%%%%%%%%%%%%%%%%%%%%%%% Appendix %%%%%%%%%%%%%%%%%%%%%%%%%%%%%%%%%%%%%%%%%%%%%%%

\appendix

\section{Tetramethylammonium hydroxide etching}\label{app:A}

In this appendix, we provide fabrication details, particularly on hard mask deposition and etching, as well as on membrane back-side alignment.

We use a SiO$_2$ back-side hard mask deposited by PECVD with which we have observed small etched holes outside of the intended membrane region. Due to their small size, the etch front terminates at a relatively shallow depth. Typically, SiO$_2$ by thermal oxidation yields a denser hard mask that is more suitable for TMAH etching. However, due to the limited thermal budget of the heterostructure (below \SI{700}{\degreeCelsius}) \cite{klos_atomistic_2024} thermal oxidation is discouraged. In spite of this, the unintentional holes neither affect the operation of the back gate nor compromise the structural integrity of the sample. 

For etching of the hard mask with buffered oxide etch, both top and back sides of the sample are coated with AZ ECI 3012 resist and hard baked at \SI{140}{\degreeCelsius} for 5 minutes to prevent diffusion of HF$^{2-}$ ions through the resist, which otherwise cause peeling and/or complete delamination of the resist. A custom sample holder mounted to a beaker lid manufactured by Advanced Micromachining Tools GmbH ensures that the top side of the sample is protected against the etchant and that the concentration of the latter remains stable throughout the process. The holder is equipped with equilibration tubes to prevent pressure buildup behind the protected surface thereby minimizing the risk of membrane rupture during etching. 

TMAH etches (001) silicon via a two-step process: (1) slow oxidation which weakens the Si-Si double back bond and (2) fast breaking of the double back bond, releasing a single Si atom into the solution in the form of Si(OH)$_4$ \cite{pal_high_2021}. The oxidation is the limiting step and depends on the amount of H$_2$O in the solution. Thus, above a concentration of about 5\%, the Si etch rate is inversely proportional to TMAH concentration, while the etch rate increases with temperature \cite{tabata_anisotropic_1992}. Moreover, the generation of H$_2$ bubbles as a byproduct of the oxidation results in a surface roughness that increases with etch rate. The Si : SiO$_2$ etch selectivity was found to increase with TMAH concentration, above 2000 : 1 for 25 \% TMAH. Given these considerations, we used 25 \% TMAH at \SI{80}{\degreeCelsius} in a 1.4 L beaker heated by a warm deionized water bath without stirring. 

\begin{figure}[h]
    \centering
    \includegraphics[width=\linewidth]{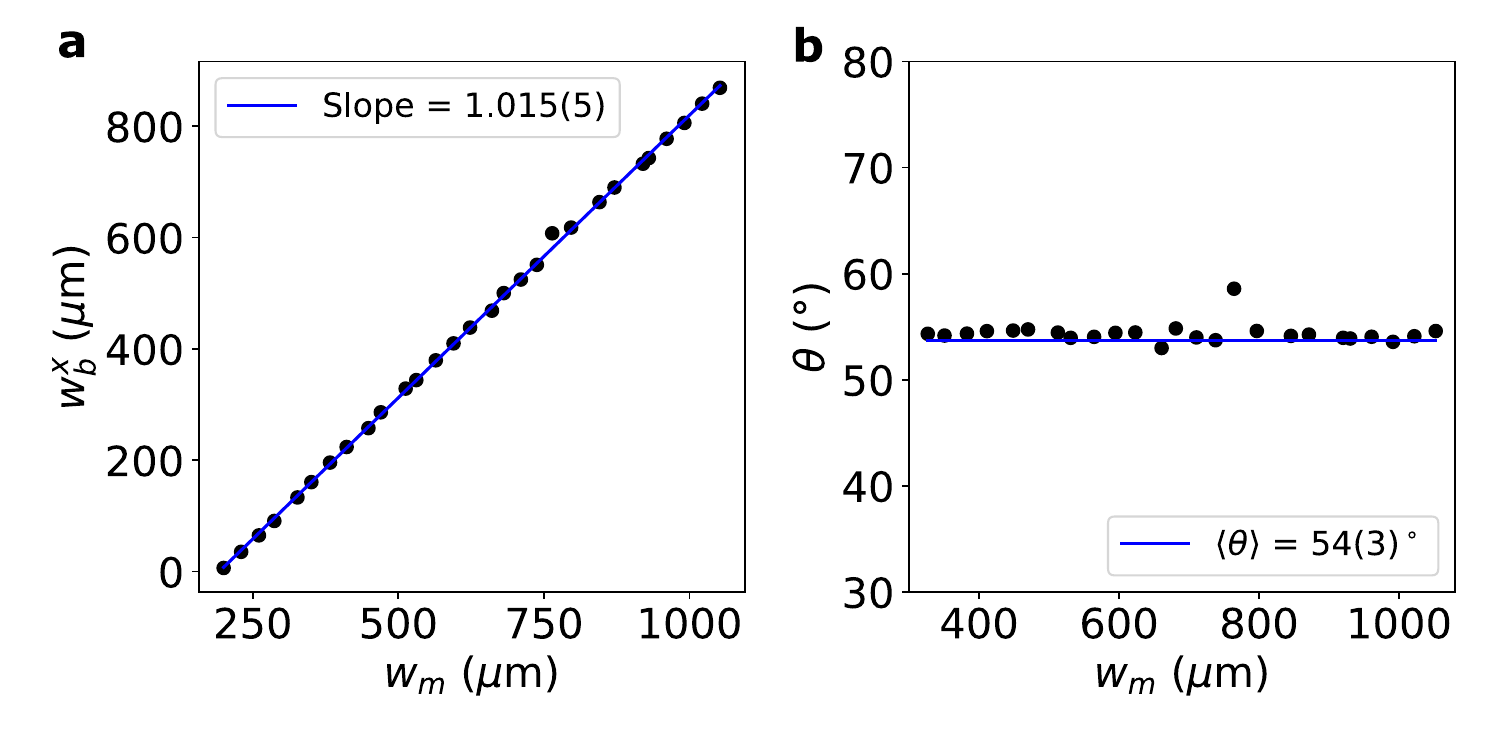}
    \caption{Wet anisotropic etching with TMAH. (a) Lateral dimensions of the membrane base $w_b$ measured along $x$ (dots) for fixed etch time $\Delta t$ and varying initial mask width $w_m$ on a reference Si sample. Linear fit to the data (blue line) to quantify control of membrane lateral dimensions. (b) Sidewall angle $\theta$ (deg) of the etch front.}
    \label{fig:8}
\end{figure}

On the same Si reference sample used to characterize ARDE in Fig.~\ref{fig:2b}, we also measured the base widths $w_b$ and computed the resulting sidewall angles $\theta = \arctan\left(\frac{2d_x}{w_m-w_b} \right)$, plotted in Fig.~\ref{fig:8}. 

\begin{figure}[h]
    \begin{center}
    \includegraphics[width=0.5\linewidth]{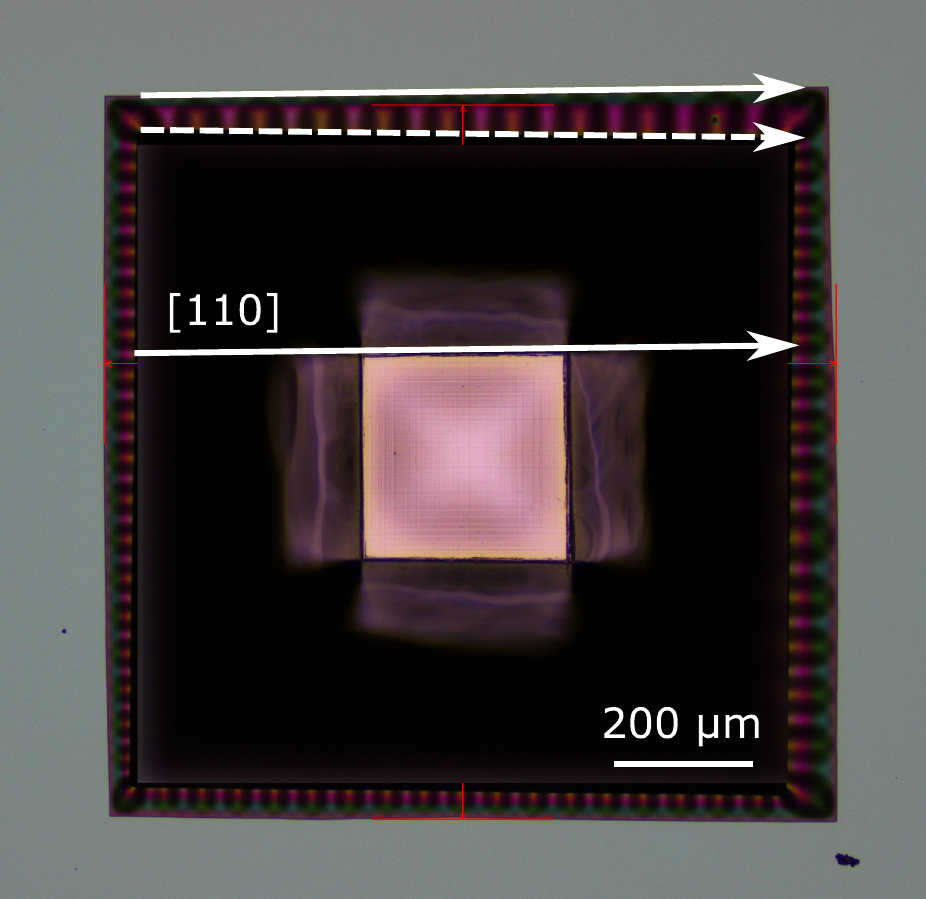}
    \end{center}
    \caption{Hard mask misalignment. Two superimposed microscope images of a membrane from the back side such that both the unetched and etched surfaces are in focus. The hard mask (dashed arrow), aligned with the diced sample edges, has a slight misalignment with $[110]$ direction (solid arrow). The etch front rotates to align itself with the crystallographic direction.}
    \label{fig:9}
\end{figure}

On some samples (especially the Si/SiGe heterostructure substrates which are cut from large wafers), the dicing misalignment leads to rotation of the etch front. For testing, membrane hard masks are aligned with the diced sample edges. For real devices, the top-side markers are also aligned with these edges, thus the misalignment propagates to the back side through BSA. Since TMAH etching on $(001)$ Si has a high selectivity to $(111)$ planes, the etch front will correct for this misalignment through a rotation (Fig.~\ref{fig:9}). Since this process is uncontrolled, it leads to asymmetries of the membrane widths $(w_b^x,w_b^y)$ and slight deviations of the membrane center with respect to that of the hard mask. This is the dominating mechanism for membrane misalignment with the top-side device, but can be mitigated by etching test markers on the back side with TMAH to calibrate the dicing misalignment. In practice, the membrane area can be made large enough to guarantee overlap with the quantum device. 

\section{Ellipsometry}\label{app:B}

\begin{figure}[h]
    \centering
    \includegraphics[width=\linewidth]{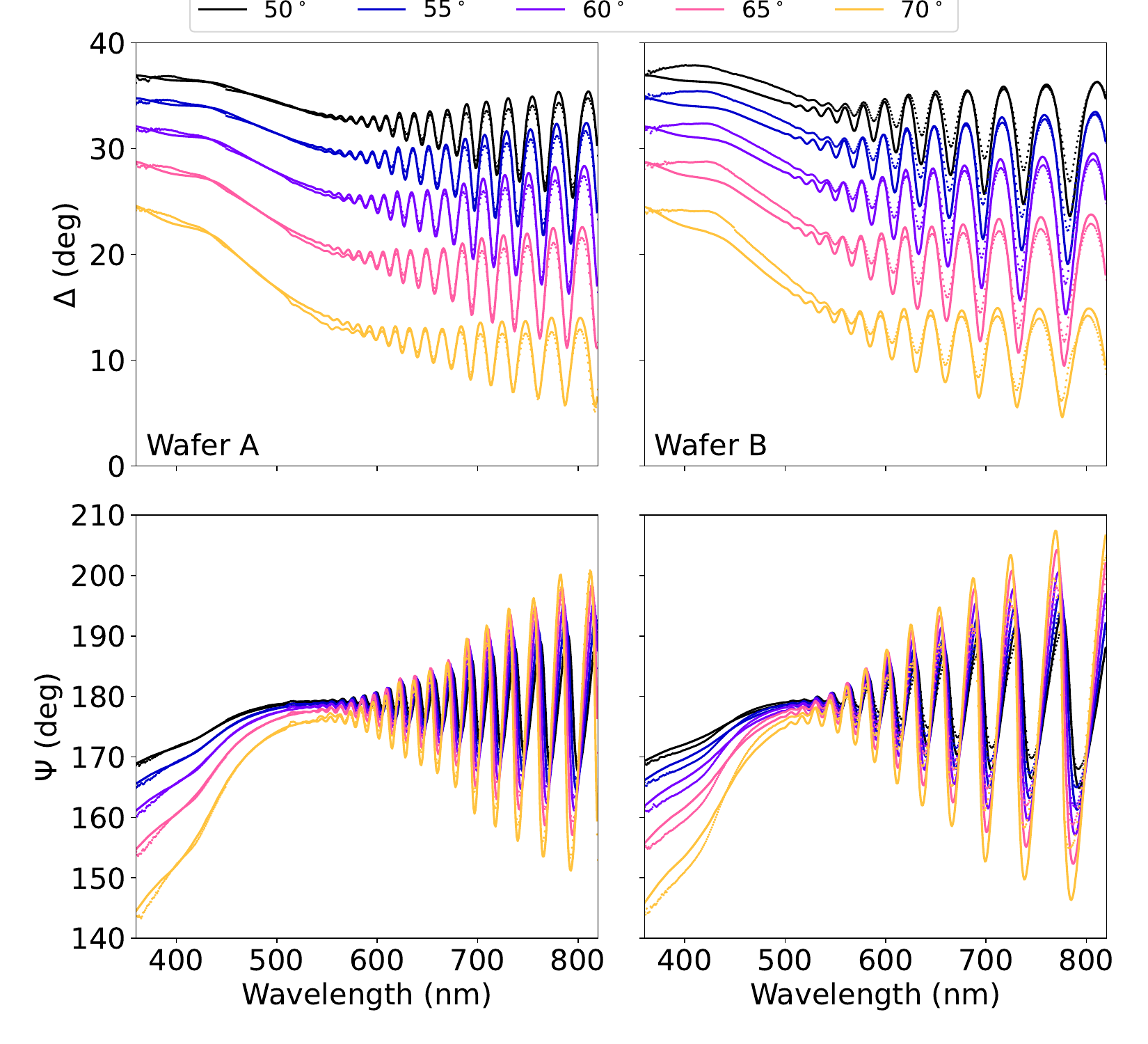}
    \caption{Ellipsometry. Measured complex reflectance ratio $\frac{r_p}{r_s} = \tan{\Psi}e^{i\Delta}$ for different incident angles and corresponding fits using the transfer matrix method on membranes 3 from wafers A and B. }
    \label{fig:10}
\end{figure}

We fit the membrane thickness $t_m$ by fitting measured ellipsometric data, namely the complex reflectance ratio between $p$- and $s$-polarizations $\frac{r_p}{r_s} = \tan{\Psi}e^{i\Delta}$ as shown in Fig.~\ref{fig:10}. The observed oscillations appear due to thin-film interference. As such, the oscillation period is the most important spectral feature for extracting the thickness. The fits agree particularly well for wafer A, while those for wafer B tend to suffer from deviations presumably due to the global buckling of membranes which might affect optical constants through strain, incident angles and incoherent effects. We use optical constants from Ref. \cite{polyanskiy_refractiveindexinfo_2024} which may deviate slightly from the sample material properties. Other possible sources of error include surface roughness and curvature of the membranes to which $\Delta$ is typically more sensitive than $\Psi$. 

\section{Membrane bending and buckling analysis}\label{app:C}

The theoretical deformations in Fig.~\ref{fig:6} were calculated by solving the Kirchhoff-Love plate equation using second-order finite differences \cite{fogang_kirchhoff-love_2021}. The generalized KL equation represents the total mechanical energy in the system including bending resistance (biharmonic terms), potential buckling energy (harmonic terms), and external load: 

\begin{align}
\mathbf{D}\cdot\nabla^4 \varphi + \textbf{N} \cdot \nabla^2 \varphi   = q
\end{align}

\noindent where $\varphi(x,y)$ is the deflection of the membrane in the $z$ direction and $q(x,y)$ is the external load applied perpendicular to the unperturbed membrane surface. The flexural rigidity (bending stiffness) $\textbf{D}$ is an extrinsic (anisotropic) property of the membrane that depends on the stiffness tensor $\textbf{C}$:

\begin{align}
    D_{ij} = \int_{-t_m/2}^{t_m/2} C_{ij} z^2 \; dz 
\end{align}

Given the symmetry of a rectangular SiGe membrane suspended at its edges and aligned in the $[110]$ $(x)$ direction, the in-plane components ($i,j=\{x,y\}$) of the flexural rigidity reduce to \cite{thomsen_silicon_2014}, 

\begin{align}
    \mathbf{D} = D_{\text{iso}}
    \begin{pmatrix}
        1 & k_2 \\
        k_2 & k_4 
    \end{pmatrix}
\end{align}

\noindent where $k_2 = 1.41$, $k_4=1$ and the isotropic flexural rigidity can be obtained from the effective compliance tensor (after in-plane $\pi/4$ rotation) $S'_{ij}$ as $D_{\text{iso}}=\frac{t_m^3}{12}\frac{S'_{22}}{S'_{11}S'_{22}-{S'}_{12}^2}$. The stress resultant $\textbf{N}$, with units of force per unit length, describes in-plane internal/residual stresses $\sigma_{ij}$ in the membrane such as epitaxial strain and is given by

Stress resultant (force per unit length):
\begin{align}
    N_{ij} = \int_{-t_m/2}^{t_m/2} \sigma_{ij} \; dz
\end{align}

\noindent Thus, the plate equation is written as

\begin{multline}
        D_{\text{iso}}\left(\frac{\partial^4\varphi}{\partial x^4} + 2k_2\frac{\partial^4 \varphi}{\partial x^2 \partial 
 y^2 } + k_4 \frac{\partial^4 \varphi}{\partial y^4}\right) \\ + N_{\text{biax}}\left(\frac{\partial^2 \varphi}{\partial x^2} + \frac{\partial^2 \varphi}{\partial y^2}\right) = q.\end{multline}

We compute the biaxial stress resultant due to the lattice mismatch between the Si and SiGe following \cite{roberts_elastically_2006} as

\begin{align}
    N_{\text{biax}} = \frac{\varepsilon_{\text{biax}}}{\frac{1}{M_{\text{SiGe}}t_m}+\frac{1}{M_{\text{Si}}t_{\text{QW}}}},
\end{align}

\noindent where we assume that the membrane is made up entirely of Si$_{70}$Ge$_{30}$ (a good assumption given the comparatively thin silicon layer) and that the quantum well generates a uniform (compressive) stress. The biaxial moduli $M_{\text{SiGe}}$ and $M_{\text{Si}}$ depend on the elastic constants, Young's moduli, and Poisson ratios which we obtain from \cite{levinstejn_properties_2001, wortman_youngs_1965}. Note that because the epitaxial strain is isotropic ($\varepsilon_{xx}=\varepsilon_{yy}$), $\mathbf{N}_{[100]}=\mathbf{N}_{[110]}$ such that the coordinates of the epitaxial stress need not be rotated.

The buckling threshold is reached when the buckling energy is equal to the bending resistance. To estimate this threshold analytically, we make the ansatz $\varphi(x,y) = \sum_{mn} \varphi_{mn }\sin(\frac{m\pi x}{w_b}) \sin(\frac{n\pi y}{w_b})$ and consider the lowest-order harmonic ($n=m=1$) with $q=0$,

\begin{align}
    \frac{\pi^2 D_{\text{iso}}}{w_b^2}(1+2k_2+k_4) - 2\sigma_{cr}t_m &= 0 
\end{align}

Defining the generalized anisotropic flexural rigidity as $\widetilde{D}_{\text{ani}} = \frac{D_{\text{iso}}}{t_m^3}(1+2k_2+k_4)$ and rearranging terms, we obtain Eq. \ref{eq:buckling_cond}.

%%%%%%%%%%%%%%%%%%%%%%%%%%%%%%%%%% Bibliography ############################################

% \bibliography{citations.bib}% Produces the bibliography via BibTeX. Comment out if .bbl file is copied into main text


\end{document}